\documentstyle[12pt,aasms4]{article}
\def\ocen{$\omega$~Cen}

\def\ltsim{\, {}^<_\sim \,}
\def\etal{{\it et al.}}
\def\ie{{\it i.e.}}
\def\eg{{\it e.g.}}

\def\deg{\ifmmode^\circ\else$^\circ$\fi}    

\def\hper{\ifmmode \rlap.^{h}\else $\rlap{.}^h$\fi} 
\def\sper{\ifmmode \rlap.^{s}\else $\rlap{.}^s$\fi}    

\def\hr{${}^{\rm h}$}
\def\mn{${}^{\rm m}$}

\def\Sc{${}^{\rm s}$\llap{.}}

\def\Sec{${}^{\prime\prime}$\llap{.}}
\def\deg{${}^\circ$}
\def\min{${}^{\prime}$}
\def\sec{${}^{\prime\prime}$}
\def\gtsim{ \,{}^>_\sim\, }
\def\hst{{\it HST\/}}
\def\vmi{\hbox{\it V--I\/}}

\def\ngc#1{NGC$\,$#1}
\def\margin#1{\marginpar{\hfill \scriptsize #1\ }}
\def\today{\number\year\space \ifcase\month\or
  January\or February\or March\or April\or May\or June\or
  July\or August\or September\or October\or November\or December\fi
  \space\number\day}
\def\now{\number\year\space \ifcase\month\or
  January\or February\or March\or April\or May\or June\or
  July\or August\or September\or October\or November\or December\fi
  \space\number\day .\number\time}
\reversemarginpar

\lefthead{P.~B.~Stetson}
\righthead{CTE Effects in WFPC2}

\begin{document}

\title{On the Photometric Consequences of Charge-Transfer Inefficiency in
WFPC2}

\author{Peter B. Stetson\footnote{Guest user, Canadian Astronomy
Data Centre, which is operated for the National Research Council
and the Canadian Space Agency by the Dominion Astrophysical
Observatory, part of the Herzberg Institute of Astrophysics}$^,$
\footnote{Guest user, Isaac Newton Group Archive, which until
1998 October 31 was operated by the Royal Greenwich Observatory}}

\affil{Dominion Astrophysical Observatory, Herzberg Institute of
Astrophysics, National Research Council, 5071 West Saanich Road,
Victoria, British Columbia V8X 4M6, Canada
\\Electronic mail: peter.stetson@hia.nrc.ca}

\centerline{and}

\affil{Mount Stromlo and Siding Spring Observatories,
Institute of Advanced Studies, Australian National University,
Weston Creek, ACT 2611, Australia}

\vspace{2 in}
\begin{abstract}

Charge-transfer effects in photometry with Wide Field Planetary Camera~2
aboard the {\it Hubble Space Telescope\/} are investigated by a comparison
of WFPC2 observations with groundbased photometry for the Galactic
globular clusters \ocen tauri and \ngc{2419}.  Simple numerical formulae
describing the fraction of lost signal as functions of position on the
detector, stellar brightness, and the diffuse sky brightness recorded in
an image are presented, and the resulting corrections are compared to
those previously derived by Whitmore \& Heyer (1997, {\it Instrument
Science Report WFPC2 97-08}).  Significant lost-charge effects are seen
that are proportional to both the $Y$ coordinate (\ie, the number of
shifts along the parallel register during readout) and the $X$ coordinate
(number of shifts along the serial register).  A ``typical'' star image
(one containing $\sim 10^4$ photoelectrons) near the center of a
``typical'' intermediate-length exposure (one with a diffuse sky
brightness of $\sim\,$10$\,$e$^-$/pixel, obtained at a camera temperature
of --88\deg$\,$C) loses approximately 2\% of its electrons to charge traps
during readout; a star in the corner of the image most remote from the
readout electronics loses twice that.  The percentage of charge lost
decreases as the star brightness or the diffuse sky brightness increases.
Charge losses during the brief period when WFPC2 was operated at a
temperature of --76\deg$\,$C were approximately 85\% greater, but apart
from that no significant change in the charge transfer losses with time
during the first 3.5 years of WFPC2's mission is evident, except possibly
a weak effect for the very faintest star images.  These results are quite
similar to those of Whitmore \& Heyer, which were based on a much smaller
data set, but there are some differences in detail.  Even with the present
set of corrections, additional sources of calibration uncertainty which I
am unable identify or characterize with the available data probably limit
the {\it external\/} accuracy of photometry from WFPC2 to of order
1--2\%.

\end{abstract}


\section{Background}

The CCDs used in the Wide Field Planetary Camera 2 (``WFPC2'') of the {\it
Hubble Space Telescope\/} are known to have imperfect charge-transfer
efficiency (``CTE'').  This property of the camera has adverse photometric
consequences that have proven difficult to characterize quantitatively.
The source of the problem appears to be chemical impurities in the crystal
lattice of the detectors that trap some photoelectrons during exposure and
readout, resulting in an underestimate of the original luminous flux
impingent on the camera.  These missing electrons eventually diffuse out
of the traps and contribute to the background noise of that same exposure
or a subsequent one.  Since the charge packets that represent stellar
images recorded at high row numbers must be clocked across a large
fraction of the chip during readout, it is to be expected that they will
encounter more traps and will lose more electrons than images recorded at
lower row numbers.  On the other hand, in frames where the diffuse sky
brightness is high --- such as in long exposures through broad-band
filters --- charge traps over the entire detector area can be filled by
sky-generated photoelectrons.  Under these circumstances, the net effect
of the charge-transfer inefficiency would be a small overall lowering of
the diffuse sky brightness perceived in the astronomical scene, while the
apparent projection of the stellar images {\it above\/} the diffuse sky
surface brightness could be comparatively unaffected.  An early analysis
suggested that in images where the sky brightness is low
($\ltsim$20--30$\,$e$^-$/pixel) a simple ``ramp'' correction proportional
to the $Y$-coordinate position of the star image, amounting to 4\% per 800
rows, would approximately correct recorded fluxes for the lost electrons;
in images where the sky was bright ($\gtsim$250$\,$e$^-$/pixel), no
correction was needed; and in cases of moderate sky brightness an
intermediate correction was recommended (Holtzman \etal\ 1995a).

While Holtzman's ramp model corrects for the first-order effects of charge
loss during readout, second-order effects may be significant.  For
instance, one would expect any given charge trap to absorb only a limited
number of electrons before their mutual repulsion effectively blocked the
ingestion of any additional charge.  Thus, faint star images consisting of
comparatively few photoelectrons should lose a larger fraction of their
total charge than images of bright stars containing many photoelectrons.
One would therefore expect faint stars to require a larger fractional
correction than bright stars near the same row number and in the presence
of the same sky brightness.  On the other hand, the large electron swarm
corresponding to the image of a bright star would occupy a somewhat larger
volume within the three-dimensional body of the detector, and might
therefore actually enounter more lattice imperfections than would be
accessible to smaller electron clouds.  Consequently, while we might still
expect that brighter stars will require smaller magnitude corrections than
fainter ones, it may be that the correction will approach zero more slowly
than the expected simple exponential dependence as arbitrarily bright
stars are considered.  In addition, while it has been demonstrated that a
given star's electron swarm can lose some charge while being clocked
through some number of rows along a CCD's parallel register, it is also
possible that said packet may lose yet more charge while being transferred
through some number of columns along the serial register.  Again, this
second ramp correction, presumably proportional to the star's
$X$-coordinate position in the digital image, may also depend upon the
sky brightness in the scene and on the brightness of the star image itself. 
The numerical coefficients of the $Y$ and $X$ ramp corrections may be
different, due to the dissimilar morphologies of the parallel and serial
registers within the semiconductor material.

Whitmore \& Heyer (1997; WH97) have estimated the above first- and
second-order effects of the charge-transfer inefficiency in WFPC2, based
upon a special sequence of observations of the globular cluster \ocen tauri
(= \ngc{5139}).  Observations were obtained at four pointings of the
telescope, with deliberate translations imposed so that a particular piece
of the star field could be recorded on the four different chips during the
course of the sequence.  Since the natural coordinate systems of the four
detectors are rotated by 90\deg\ with respect to each other, this allowed
several hundred star images to be used for a purely differential
determination of the fraction of charge lost as a function of position on
the detector, without a need to know the absolute brightness of any given
object.  WH97 presented their results in four alternative formulations, of
which the most relevant to the present discussion is given by their Eqs.~1,
2d, and 3d:  
$$\hbox{\it CTS}_{\hbox{\scriptsize corrected}} = \left[ 1 +
{\hbox{$Y$-CTE}\over{100}} \times{Y\over800} +
{\hbox{$X$-CTE}\over{100}}\times{X\over800}\right] \hbox{\it
CTS}_{\hbox{\scriptsize observed}}$$ 
$$Y\hbox{-CTE} =
10^{\hbox{\small\,$(0.6930-0.2675\times\log{\hbox{\it BKG}_{\hbox{\scriptsize
blank}}})$}}$$ 
$$X\hbox{-CTE} = 7.040 - 1.63 \times \log{CTS_{\hbox{\scriptsize
observed}}}$$ 
Here, {\it CTS\/} represents total WFPC2 data numbers belonging to the
stellar image within an aperture 0\Sec5 in radius, as perceived with the
gain = 7$\,$e$^-$/DN electronics; $X$ and $Y$ represent the position of the
stellar image in the natural coordinate system of the detector (pixel (1,1)
being the first one to be read through the amplifier and (800,800) being
the last, with the $X$ coordinate varying more rapidly); and $\hbox{\it
BKG}_{\hbox{\scriptsize blank}}$ is the perceived diffuse sky brightness in the
scene, exclusive of recognized stellar images.  (I have opportunistically
adopted this version from the four variants presented, because ---
following Holtzman \etal\ (1995b) --- in reducing WFPC2 data I routinely
correct measured magnitudes to their half-arcsecond-aperture equivalents,
and because the data files produced by my software contain an estimate of
the local sky brightness around and underlying each detected object.)

Certain properties of these formulae may be noted.  

(1)~The correction that is proportional to the $Y$ position of the star
depends only on the diffuse sky brightness in the image, and not on the
brightness of the star itself.  In contrast, the alternative $Y$ ramp
corrections presented for magnitudes as measured through 0\Sec2 synthetic
apertures depend upon both sky and star brightness.  This change in
formulation stems from the finding by WH97 that the difference between
magnitudes measured in 0\Sec2 and 0\Sec5 apertures itself depends upon
the stellar brightness: ``We find the CTE loss is slightly reduced wthen
the larger aperture is used, by about 1\% for the bright stars and 3\% for
the faint stars.''  Apparently, in the data studied by WH97 this effect
tended to erase the magnitude dependence of the ramp corrections in the
larger aperture.  However, it should be noted that a simple differencing
of the formulae provided by WH97 indicates that perfect cancellation can
occur only at a particular value of the sky brightness, and will not be
the case in general.

(2)~The correction that is proportional to the $X$ coordinate depends only
on the brightness of the star and not at all on the diffuse brightness of
the sky, a result not necessarily to be predicted from the electron-trap
model of the charge-transfer inefficiency.

(3)~The $Y$-coordinate ramp is proportional to an exponential function of
the logarithm of the sky brightness, so (a)~the correction tends
asymptotically (but slowly) to zero as the sky brightness increases
without limit, but (b)~as the sky brightness decreases to zero, the
correction
$$10^{\hbox{\small $\,0.6930$}}\times10^{\hbox{\small $-0.2675\times\log{\hbox{\it
BKG}_{\hbox{\scriptsize blank}}}$}} = \hbox{\it constant}\,\times\,{1\over
\hbox{\it BKG}^{0.2675}_{\hbox{\scriptsize blank}}} \rightarrow \infty.$$ 
This implies a non-physical result for a star of any apparent magnitude
detected in the presence of a very low diffuse sky brightness.

(4)~The $X$-coordinate ramp is a linear function of the logarithm of the
star counts, so (a)~the correction again diverges as the
number of star counts approaches zero.  However, in this case the
divergence is not a serious drawback, since photometry of any stellar
image containing that few photoelectrons is unlikely to be of genuine
astronomical interest.  Conversely, (b)~the correction is identically zero
when $\log{CTS_{\hbox{\scriptsize observed}}} = 7.040/1.63 = 4.32$, and changes
sign (\ie, the lattice imperfections in the serial register appear to act
as electron sources rather than electron sinks) for stellar images
containing more than 21,000~DN within a half-arcsecond aperture.  This is
approximately the amount of flux contained within a star image whose
central pixel just equals the digital saturation level of the
analog-to-digital converters in WFPC2 when the gain$\,=\,7\,$e$^-$/DN
electronics are used.  Therefore, this extrapolation, too, is probably of
minimal practical relevance except possibly for very bright stars observed
with the high-gain electronics bay.  These extrapolations do emphasize,
however, that the WH97 corrections are purely empirical constructs, and
are not tied directly to a self-consistent physical model of the nature of
the charge-transfer inefficiency.

(5)~The model correction makes no allowance for the possibility of a
charge loss that is independent of position, regardless of whatever
dependence it may have on either stellar brightness or sky flux.  This
feature of the model is a direct consequence of the experimental design,
where apparent flux ratios were observed for individual stars recorded at
various positions on the chips, without any consideration of (a)~flux
ratios for different stars with externally known magnitude differences, or
(b)~flux ratios for images of individual stars as recorded in frames with
different exposure times.  Charge losses that are independent of position
may be expected within the lattice-flaw model:  simply by virtue of being
formed in the body of the silicon wafer in the first place, the stellar
image may lose some charge to traps located near the place of its
formation.  In addition, any inefficiency inherent in the transfer of
charge packets from the top of a parallel register to the serial register,
or from the end of the serial register into the on-chip amplifier, would
be perceived as a charge loss that is independent of the original location
of the stellar image on the detector.

A hint that position-independent charge losses {\it might\/} be
significant came from preliminary analyses of WFPC2 images of the
globular clusters \ngc{2419} and Palomar~4 (see W.~E.~Harris \etal\ 1997;
Stetson \etal\ 1998, 1999).  A comparison of short-exposure
($\leq 60$~s) images to ground-based photometry of the same objects
implied photometric zero-points consistent with those found by Holtzman
\etal\ (1995b) from similarly short exposures of the \ocen\ standard field.
In contrast, comparison of longer ($\geq 1200$~s) exposures of the same
fields to the same ground-based photometry implied zero points that were
different by approximately 0.05~mag, in the sense that the effective
quantum efficiency of the detectors was higher in the longer exposures.  A
subsequent recalibration of the ground-based photometry, including the
addition of many more independent ground-based and HST observations of
\ngc{2419}, suggested that the original estimate of 0.05~mag might in
fact have been high; the additional data suggested that an anomaly of
order 0.03~mag on average might be more nearly correct.  With the
publication of the WH97 corrections, it is apparent that the perceived
short-minus-long zero-point difference might just be a reflection of the
dependence of the fractional charge loss on the total number of recorded
stellar photoelectrons and on the perceived surface flux of the sky.
However, it remains possible that some residual position-independent
charge loss could occur.

In this paper I reinvestigate the photometric consequences of the
charge-transfer inefficiency in WFPC2, based on extensive series of
observations of the Galactic globular clusters \ocen tauri and
\ngc{2419}.  In addition to testing the validity of the corrections
proposed by WH97 on a much larger data set, I investigate certain
aspects of the problem not considered by them.

\section{A New Analysis of Charge Loss Effects --- Overview}

The data analyzed in this paper are all in the public domain, and were
acquired from the {\it HST\/} Science Archive of the Canadian Astronomy
Data Centre.  The observations include 318 individual WFPC2 exposures
lying within 2~arcminutes of the nominal center of the \hst\ standard
calibration field in the nearby Galactic globular \ocen tauri (Equinox
2000.0 coordinates 13\hr25\mn37\Sc0, --47\deg35\min38\sec\ according to
H.~C.~Harris \etal\ 1993).  To these are added 91 exposures mostly centered
within a few dozen arcseconds of 7\hr38\mn06\Sc8, +38\deg51\min55\sec:
about 1\min\ south of, but including, the center of the far-halo globular
cluster \ngc{2419}.  These latter data include all of the observations
discussed by W.~E.~Harris \etal\ (1997), plus some additional exposures of
the cluster that were obtained by STScI staff specifically to assist in
the calibration of WFPC2.  Some differences between the present data set
and that of Whitmore \& Heyer (1997) deserve particular mention.

(1)~The WH97 analysis relied on eight observations made in each of the
standard broad-band filters (F336W, F439W, F555W, F675W, and F814W), as
well as four observations in F606W intended to allow a comparison with a
skyflat developed during the Medium Deep Survey project.  From each of the
44 exposures, data from only one of the four WFPC2 chips were
considered in their analysis.  The present study concentrates on
observations made in the F555W and F814W filters (\ocen:  152 and 159
exposures, respectively; \ngc{2419}: 27 and 58 exposures), because these
are particularly relevant for the \hst\/ {\it Key Project on the
Extragalactic Distance
Scale\/}\footnote{http://www.ipac.caltech.edu/H0kp/} and for studies of
certain Local Group star clusters and galaxies in which I am especially
interested.  Data from all four chips of each of the 402 $V$ and $I$
exposures were considered in the analysis below.  Seven observations of
\ocen\ in the intermediate-band F547M filter were also included in the
reductions, to allow the eventual estimation of the fundamental photometric
zero-points for this filter for a particular project, but these data were
not utilized in the analysis of the charge-transfer inefficiency to be
presented below.  

(2)~The WH97 data were all taken on 1996~June~29, whereas the \ocen\ data
employed here were taken from 1994~January~11 through 1997~June~26.  The
\ngc{2419} data were all obtained during either 1994~May~21--22 (the 28
W.~E.~Harris observations), 1996~December~21 (6 exposures), or
1997~November~18--19 (57 exposures).  Since the observation date 
determines the spacecraft roll angle required to direct the
solar panels toward the Sun while the telescope is pointed at a particular
spot on the celestial sphere, the WH97 observations were all obtained at
the same telescope orientation.  The \ocen\ observations discussed here,
having been obtained at essentially all times of the year, represent
several full cycles of spacecraft roll angle.  The \ngc{2419} observations
were also taken at several different orientations.

(3)~The WH97 data were all taken with WFPC2 operating at a camera
temperature of --88\deg~C, whereas 116 of the 318 \ocen\ observations
considered here were made at the warmer temperature of --76\deg~C used
early in WFPC2's mission; the remaining 202 \ocen\ observations and all 91
\ngc{2419} exposures were taken at --88\deg~C.

(4)~The WH97 data were obtained with Electronics Bay 4, which gives a
nominal gain of 7$\,$e$^-$/DN, whereas 264 of the \ocen\ observations
analysed here were obtained through Electronics Bay 3 (gain approximately
14$\,$e$^-$/DN).  The rest of the \ocen\ and all the \ngc{2419} observations
were taken at 7$\,$e$^-$/DN.

(5)~The WH97 analysis was purely differential on a star-by-star basis,
considering only the flux {\it ratio\/} between two observations of a
given star as a function of the positional difference between the two
stellar images in the natural coordinate systems of the detectors.  The
new analysis will assume that the true standard-system $V$ (Johnson) and
$I$ (Kron-Cousins) magnitudes are known {\it a priori\/} from ground-based
observations of the two target fields.  The transformation equations
relating the \hst\ natural-system magnitudes to the groundbased standard
system will be determined by a robust least-squares technique, assuming
that the color transformation coefficients of Holtzman \etal\ (1995b;
their Table 7) are correct.  The statistical solution of these equations
will yield the fundamental zero points of the four WFPC2 CCDs, and the
residuals of individual stars from these solutions will serve as
diagnostics of the charge-transfer inefficiency.

(6)~WH97 estimated the mean clear-sky surface brightness for a given frame
from a star-free region near the center of the field; this value was then
considered appropriate for every star in the image.  The software used in
the present analysis estimates a unique local sky-brightness value for
every star, from the mode of brightness values in a surrounding annulus if
it is aperture photometry that is being derived, or from the median
brightness value at and around the position of the star in a residual
image from which all known stars have been subtracted in the
profile-fitting analysis.  Because these individual, local brightness
measurements are conveniently available, they will be used in the
following analyses rather than a global average sky for the frame as a
whole.  For the globular-cluster images used here, there is little net
difference between the two methods (typically $\sim 0.04\,$e$^-$/pixel).

Both the present study and that of WH97 include a few observations
obtained with internal preflashes, whose purpose was to increase the
leverage for determining the dependence of the charge loss on diffuse sky
brightness:  WH97 had three exposures in the F555W filter with some
preflash, while the present sample includes those three plus another nine
preflashed exposures of \ngc{2419}.  In each case, the preflash produced a
diffuse surface brightness averaging approximately 175$\,$e$^-$/pixel.  In
the long exposures of \ngc{2419} the sky brightness was typically of order
250$\,$e$^-$/pixel, and in all other cases the background level was quite
close to zero.

All WFPC2 images studied here were recalibrated by the Canadian Astronomy
Data Centre, using the best bias and flat-field frames available at the
time the data were requested.  The intensities in the frames were
corrected for geometric distortion, known defective pixels were masked,
and the resulting images were multiplied by 4.0 and coverted to short
integers as described by Stetson \etal\ (1998).  Instrumental magnitudes
were determined via the profile-fitting technique with the computer
program ALLFRAME (Stetson 1994), and selected bright, isolated stars were
used to compute magnitude corrections to a system defined by
half-arcsecond synthetic apertures via a growth-curve analysis (Stetson
1990).  All transformations from the WFPC2 instrumental system to the
standard ground-based $V,I$ system were performed with a version of the
program CCDSTD (see Stetson 1993).  Because the edges of the pyramid
mirror in WFPC2 are imaged onto the photosensitive areas of the four
camera CCDs, this analysis will be confined to that area of each chip
which is almost certainly unvignetted; this has been taken to be $75 < X,Y
< 800$ in the three WFC chips, and $100 < X,Y < 800$ in PC.

The ground-based Johnson $V$ and Kron-Cousins $I$ magnitudes which will be
used for calibrating WFPC2 are taken from Walker (1994) for \ocen, and
from my own analysis of ground-based data for \ngc{2419}.  The photometry
for the latter cluster is based upon 68 $V$ and 62 $I$ frames obtained
during the course of 14 nights in six observing runs.  The images were
taken with the Kitt Peak 4m and 2.1m telescopes, the
Canada-France-Hawaii Telescope (data obtained through the courtesy of the
Canadian Astronomy Data Centre), and the Isaac Newton Telescope (data
obtained through the courtesy of the Isaac Newton Group Archive), and have
been rigorously transformed to the photometric system of Landolt (1992) as
part of an ongoing program that has now homogenized photometric data from
over 40 individual observing runs.  

\ngc{2419} is a massive, rich globular cluster located about 100~kpc away,
and the field is quite crowded under groundbased seeing conditions.
Therefore the HST imagery was examined to identify stars that appear to be
minimally crowded.  Wherever possible, a star identified in the
groundbased images was cross-identified with a star detected in the WFPC2
data.  Then the groundbased photometry was differentially corrected for
those neighbors that were found in the HST images but were {\it not\/}
identified and accounted for in the analysis of the groundbased data.  Any
star that required a blending correction in excess of 0.1~mag was rejected
as a potential standard.  As the analysis proceeded, any star in either
\ocen\ or \ngc{2419} for which the mean of the calibrated HST photometry
differed from the mean of the groundbased data by more than 0.3~mag in
either filter was likewise discarded, and the solutions were redone until
no further rejections were required.  \ocen\ stars that were rejected at
this stage were Walker numbers 1, 9, 24, 48, 49, 57, 60, 77, 99, 129, 139,
152, 154, 174, 196, 203, 238, 240, 243, 247, 248, 261, 274, and 275; this
left 184 \ocen\ stars for which the groundbased and HST photometry were
concordant to within $\pm 0.3$~mag.  Fifteen of 366 stars in \ngc{2419}
were similarly discarded.

It may be argued that the inclusion of stars with such large standard
errors would vitiate any delicate photometric analysis.  However, this
should not be a serious problem.  In the analyses that follow all
observations are weighted in accordance with their inferred standard
errors, which are fairly well known from the photon statistics and
repeatability from multiple images.  Then a robust version of least
squares is used (see Stetson 1989) which permits the solution to follow
the main trend of the data without being unduly influenced by outliers.
Inclusion of faint stars is necessary to provide the needed leverage on
the magnitude dependence of the charge loss, and a more strict censoring
of the data might introduce bias rather than reduce it.  In addition, most
stars were eventually imaged at many different locations on the various
chips.  A star whose photometric errors give a semblance of a large charge
loss when imaged at a high value of $X$ or $Y$ would produce a cancelling
effect on other occasions when it appeared at low column or row numbers.
As in the differential WH97 study, then, gradients in the photometric zero
point across the face of the chips, and variations of those gradients with
sky brightness and stellar magnitude will not be sensitive to the accuracy
of the groundbased magnitudes.  Prior knowledge of the standard
photometric magnitudes will be useful only in testing for the presence of
position-independent photometric non-linearities, and those tests alone
will be affected by uncertainty in the groundbased results.

The present data set will be analysed in three ways: (1)~assuming that
charge-transfer losses are negligible; (2)~correcting the observed
stellar fluxes according to the precepts of WH97; and (3)~solving for a
new numerical model of the charge loss as a function of position, stellar
brightness, and sky brightness simultaneously with the determination of
the photometric zero points.  The results of these three analyses will
be presented in \S\S3.1, .2, and .3, respectively, for the data obtained
at a camera temperature of --88\deg$,$C.  A briefer discussion of the data
obtained at the warmer temperature will be presented in \S4.

The fundamental assumptions of the present analysis are:
\begin{enumerate} \item The charge-transfer inefficiency affects the
gain~=~7 and the gain~=~14 observations identically, provided
care is taken to evaluate the charge loss in units of electrons.  \item
The inefficiency is the same for F555W and F814W.  \item The size of the
effect has not evolved significantly between 1994~May and 1997~November.
\item The charge loss may be significantly different at a camera
temperature of --76\deg~C as opposed to --88\deg~C.  \end{enumerate}
WH97 did not require assumptions (1), (3), or (4) because all their
observations were obtained on a single date with a single gain setting and
camera temperature.  Our assumption (1) follows from the hypothesis that
the charge transfer inefficiency is completely the result of physical
processes within the body of the CCD, and not from subsequent effects that
take place during amplification and digitization.  Assumptions (2) and (3)
are made primarily for simplicity and ease of modifying existing software
to perform the present analysis, but the work of WH97 is relevant for
assessing the validity of assumption (2), at least.  While they found that
the gross correction for lost charge does appear to depend upon the filter
--- the corrections being the greatest at short and long wavelengths and
being least for the F555W and F675W filters --- they noted, ``This is
probably because the chips are more efficient at these wavelengths [\ie,
near 555 and 675$\,$nm], hence the background is higher.  The higher
background appears to reduce the CTE loss\ldots'' Thus, when proper
quantitative account is taken of the apparent dependence of charge loss on
background brightness level, WH97 suggest that the photometric correction
does not depend heavily on bandpass.  In any case, the validity of
assumptions (2) and (3) can be checked {\it ex post facto\/} once the
average baseline effect of charge-transfer inefficiency has been
calibrated and removed.  As a result of assumption (4), I have performed
separate and independent analyses for the colder and warmer camera
temperatures.

\section{A New Analysis of Charge Loss Effects --- Computations}

\subsection{No corrections}

A least-squares fit of the basic transformation equations to the images of
36,234 stars recorded in the cold-camera observations of \ocen\ and
\ngc{2419} yielded the photometric zero points listed in
Table~1.\margin{Table~1}  As mentioned above, the color coefficients
derived by Holtzman \etal\ (1995b) were imposed as givens.  Specifically, 
\begin{eqnarray}
F555W = V + A_V + 0.052\times(V-I) - 0.027\times(V-I)^2 \nonumber \\
F814W = I + A_I + 0.062\times(V-I) - 0.025\times(V-I)^2. \nonumber
\end{eqnarray}
The first part of the table gives the zero point for each chip/filter
combination based upon the full set of observations obtained with
electronics bay 3 (exclusively short exposures of the \ocen\ field), and
the full set with electronics bay 4 (a combination of short exposures of
\ocen\ and both short and long exposures of \ngc{2419}).  In the second
part of the table, electronics bay 4 zero points are presented for the
short \ocen\ exposures, the short \ngc{2419} exposures, and the long
\ngc{2419} exposures considered separately.  All of these derived zero
points are tabulated in detail because they will be used below to test the
validity of the various correction schemes.

The calibration residuals $\delta$(magnitude) resulting from the 16
combined solutions (2~gain settings~$\times$ 2~filters~$\times$ 4~chips,
corresponding to the zero points listed in Table~1a) were all merged
into a single file; the sense of the residual is (observed instrumental
magnitude) {\it minus\/} (magnitude predicted from the standard-system
indices and the mean transformation equations).  Figure~1 \margin{Fig.~1}
shows the individual residuals plotted versus (top)~$X$ coordinate on the
chip, (middle)~$Y$ coordinate on the chip, and (bottom)~raw instrumental
magnitude $m$ (= {\it constant\/} -- $2.5 \log$[number of stellar
photoelectrons within a radius of 0\Sec5]); for clarity, only unsaturated
stars where the combined uncertainty of the ground-based standard magnitude
and the WFPC2 instrumental magnitude is less than 0.05~mag have been
plotted.  It is apparent that there are significant residual trends with
both the $Y$ and $X$ coordinates, but no systematic nonlinearity with
magnitude is obvious.  To provide a more quantitative, albeit still crude,
measure of the magnitude of these effects, a robust, weighted least-squares
surface of the form
$$\delta = a + {\partial \delta\over \partial X}\cdot X + 
{\partial \delta \over \partial Y}\cdot Y + 
{\partial \delta \over \partial m}\cdot m +
{\partial^2 \delta \over \partial X \partial m}\cdot X\cdot m +
{\partial^2 \delta \over \partial Y \partial m}\cdot Y\cdot m +
{\partial \delta \over \partial T}$$
was fitted to all 36,234 fitting residuals.  The natural-system
position variables are $X$ and $Y$, $m$ is the instrumental magnitude
as defined above, and $T$ is the date of observation, measured in
years from Heliocentric Julian Date 2,450,000.0.  To place the amplitude of
the position and magnitude trends in a comfortable set of units, I take
the maximum displacement of any given well-measured star from the mean of
all stars to be $\left|\Delta X\right| < 400$~px, $\left|\Delta Y\right| <
400$~px, $\left|\Delta m\right| < 2.5$~mag.  From the least-squares fit I
estimate the maximum photometric consequences of the neglected charge-loss
corrections to be 
\begin{eqnarray}
{\partial\delta\over\partial X}\cdot(400\hbox{\ px}) & = & +0.0150\pm 0.0006\ \hbox{mag}, \nonumber \\
{\partial\delta\over\partial Y}\cdot(400\hbox{\ px}) & = & +0.0154\pm 0.0006\ \hbox{mag}, \nonumber \\
{\partial\delta\over\partial m}\cdot(2.5\hbox{\ mag}) & = & +0.0142\pm 0.0004\ \hbox{mag}, \nonumber \\
{\partial^2\delta\over\partial X\partial m}\cdot(400\hbox{\ px})\cdot(2.5\hbox{\ mag}) & = & +0.0038\pm 0.0008\ \hbox{mag}, \nonumber \\
{\partial^2\delta\over\partial Y\partial m}\cdot(400\hbox{\ px})\cdot(2.5\hbox{\ mag}) & = & +0.0150\pm 0.0008\ \hbox{mag}. \nonumber
\end{eqnarray}
The positive signs of the first three coefficients, representing the
linear dependence on position and magnitude, indicate that the instrumental
magnitude is too large, \ie, the star is measured too faint with respect
to the average zero-point for all stars taken together, when the
$X$ coordinate is large, the $Y$ coordinate is large, or the star is faint
(large $m$) --- all in accordance with the simple predictions of the
charge-loss model.  The positive signs of both second derivatives
demonstrate that the effect of charge loss is further compounded for faint
stars at large $X$ and $Y$ positions.  For a star in a typical exposure,
any one of these neglected corrections can cause a systematic error of no
more than about 0.015~mag, which corresponds to a root-mean-square error
of about 0.01~mag.  A star in the wrong corner of position-magnitude
space could, however, be subject to an error more than four times larger.  Note
in particular that the result for ${\partial \delta\over\partial m} =
(+5.69\pm 0.17)\times 10^{-3}$ implies a stretching of the instrumental
magnitude scale, on average:  faint stars are measured too faint compared
to bright ones, by about 0.03~mag per 5~magnitudes (an approximate
figure for the effective dynamic range of WFPC2).  

The remaining coefficient determined from the least-squares fit to the
residuals is
$${\partial \delta \over \partial T} = +0.0004 \pm 0.0002\,\hbox{mag\ yr$^{-1}$}.$$
The positive sign of the result indicates that for fixed true-system
magnitudes, the observed instrumental magnitudes are becoming larger,
\ie, fainter, as time passes.  However, with an indicated rate
$< 0.001$~mag~yr$^{-1}$, this secular change is not very important even
over the more than 3.5 years over which the cold-camera data were
obtained.
\subsection{The WH97 corrections}

In applying the WH97 corrections to the present data sets, all star and
sky fluxes were converted to units of electrons, assuming gain values of
identically 7.0 or 14.0$\,$e$^-$/DN, to facilitate combining data obtained
with either of the two electronics bays.  Since I will be calibrating
small, differential magnitude corrections directly from stellar
observations, minor differences between the postulated and actual gain
factors will not seriously damage the analysis.  The numerical
coefficients of the WH97 equations were adjusted to represent fluxes
expressed in units of electrons rather than electronics bay 4 data
numbers.

As noted above, the logarithm of the sky brightness varies rapidly as the
sky brightness approaches zero, and is undefined for any non-positive
estimate of the sky flux.  In fact, Table~1 of WH97 indicates that seven of
their 44 digital images had inferred median sky-brightness values,
$\hbox{\it BKG}_{\hbox{\scriptsize blank}}$, that were negative; a further 18
images had positive median brightness values that were less than 
0.5~DN~px$^{-1}$, where a small inaccuracy in $\hbox{\it
BKG}_{\hbox{\scriptsize blank}}$ begins to lead to a significant change in
$\log{\hbox{\it BKG}_{\hbox{\scriptsize blank}}}$. WH97 make no explicit mention
of how images like these are to be treated in their methodology.  The
electron-trap model of the charge-transfer inefficiency predicts that a
realistic stellar image would lose only a finite number of electrons even
in the limit of a hypothetical exposure with a zero effective background (a
short exposure of bright stars through a narrow filter, for instance).
Therefore, it is apparent that some modification must be made to the
formulae so that the predicted photometric correction approaches a finite
constant value as the estimated background brightness reaches (and, in the
presence of noise, possibly passes through) zero.  In the present analysis
I do this by simply ``softening'' the inferred sky brightness at low values:
\begin{eqnarray}
\hbox{\it BKG}_{\hbox{\scriptsize adopted}} & = & \left\{(1\,\hbox{e}^-)^2+
\left[\hbox{max}\left(0,\hbox{\it BKG}_{\hbox{\scriptsize observed}}\right)
\right]^2\right\}^{1\over2}.
\end{eqnarray}
(Please note that in the present analysis, the units of {\it BKG\/} are
electrons, not DN as in the WH97 study.) The softening constant,
1$\,$e$^-$, has been adopted arbitrarily, and has as its sole justification
the fact that it is better than doing nothing. Fig.~9 of WH97 indicates
that the fraction of charge lost from a stellar image continues to increase
down to $\log\left[\hbox{\it BKG\/}(DN)\right] \sim -1,$ or $\hbox{\it BKG}
\sim 0.7\,$e$^-$/pixel, but diffuse sky fluxes below this have not been
sampled. The imposition of this minimum inferred sky brightness means that
all frames where the diffuse sky brightness is small compared to
1$\,$e$^-$/pixel will be subject to similar, finite, photometric corrections. 
Henceforth in this paper, the unqualified symbol {\it BKG\/} will refer to
$\hbox{\it BKG}_{\hbox{\scriptsize adopted}}$ as defined in Eq.~1.

Having made this one modification to the WH97 photometric corrections,
I then recomputed the cold-camera zero points for the various combinations
of gain factor, chip, and filter, arriving at the results shown in
Table~2.\margin{Table~2}  I forgo plotting the residuals from these
solutions, but the numerical correlations with the various independent
variables are:
\begin{eqnarray}
{\partial\delta\over\partial X}\cdot(400\hbox{\ px}) & = & +0.0109\pm 0.0006\ \hbox{mag} \nonumber \\
{\partial\delta\over\partial Y}\cdot(400\hbox{\ px}) & = & -0.0017\pm 0.0006\ \hbox{mag} \nonumber \\
{\partial\delta\over\partial m}\cdot(2.5\hbox{\ mag}) & = & +0.0026\pm 0.0004\ \hbox{mag} \nonumber \\
{\partial^2\delta\over\partial X\partial m}\cdot(400\hbox{\ px})\cdot(2.5\hbox{\ mag}) & = & -0.0044\pm 0.0008\ \hbox{mag} \nonumber \\
{\partial^2\delta\over\partial Y\partial m}\cdot(400\hbox{\ px})\cdot(2.5\hbox{\ mag}) & = & +0.0109\pm 0.0008\ \hbox{mag} \nonumber \\
{\partial \delta\over \partial T} & = & +0.0005 \pm 0.0002\ \hbox{mag\ yr}^{-1} \nonumber 
\end{eqnarray}
Thus, the WH97 corrections operate in the right direction to reduce the
size of the maximum systematic error induced by the electron traps, but
they do not eliminate the errors entirely, at least for the present data
set:  some of the charge-loss effects are slightly over-corrected, others
are significantly under-corrected.  Most notably, in these data the amount
of charge lost in the serial register (the $X$ direction) appears to be
larger than estimated by WH97, as is the dependence of the $Y$-coordinate
ramp on instrumental magnitude (which has been set to zero in their model). 
As a result, a maximum systematic error in excess of 0.02~mag is still
possible for stars in an unfortunate location in $(X,Y,m)$ space in an
average exposure.
\subsection{New model corrections}

For internal consistency, unlike WH97 I adopt symmetric,
exponential-of-a-logarithm, formulations for the charge-transfer losses in
both $X$ and $Y$ as functions of both star and sky flux.  In doing this,
I claim no superior knowledge of the detailed physics of charge loss, but
rather follow WH97's lead in postulating a simple, phenomenological,
numerical model for the required photometric corrections.  At the same
time, in reconsidering the WH97 equations, I have tried to improve on
some of the aforementioned features of their formulation.

In particular, I have added a term to the formula which allows for the
possibility of a charge loss which is independent of the image's position
on the chip.  In doing this, I have arbitrarily assumed that any charge
loss which {\it is\/} independent of position will have precisely the
same dependence on sky brightness and stellar flux as the $Y$-dependent
ramp correction.  Again, this assumption is arbitrary and has been made to
reduce the number of independent coefficients that must be determined; it
is physically reasonable provided that the electron traps contributing to
any position-independent charge loss are physically similar to those that
produce the well-documented $Y$ photometric ramp in the body of the
detector, which may be different from the traps that produce the
less well-understood $X$-dependent charge losses that probably occur in
the serial register.  Of course, if it should happen that there are no
position-independent effects, the least-squares analysis will allow the
coefficient of the constant term to assume a negligibly small value.

Finally, rather than expressing the equations in terms of the image
positions and the star and sky brightness in absolute terms, I consider
expansions about ``typical'' values of these variables.  This makes the
various terms in the formulae more nearly independent of each other, so
that the least-squares solution is less ill-determined, and also leads to
derived standard errors of the coefficients that can be understood in a
simple and direct way.

With {\it BKG\/} representing the softened sky flux in e$^-$/pixel,
{\it CTS\/} representing the stellar flux within a 0\Sec5 (radius)
aperture in e$^-$, and $X$ and $Y$ the position of the star image in the
natural coordinate system of the chip, the adopted numerical model for the
photometric corrections is: 
\begin{eqnarray}
\Delta X & = & {{X-425}\over 375}; \\
\qquad \Delta Y & = & {{Y-425}\over 375}; \\
\hbox{\it sky}  & = & \log_{10}{[\hbox{\it BKG}}] - 1; \\
\hbox{\it star} & = & \log_{10}{[\hbox{\it CTS}}] - 4; \\
Y\hbox{-CTE} & = & c_1 + c_2\times\hbox{\it sky} + c_3\times\hbox{\it star}; \\
X\hbox{-CTE} & = & c_4 + c_5\times\hbox{\it sky} + c_6\times\hbox{\it star}; \\
C & = & 1 + 0.01\times\left(c_7 + \Delta Y\right)\times e^{\hbox{\small
$Y$-CTE}} + 0.01\times \Delta X\times e^{\hbox{\small $X$-CTE}}.
\end{eqnarray}
If $m$ represents the measured stellar magnitude relative to whatever
zero point, then
\begin{eqnarray}
m(\hbox{corrected}) = m(\hbox{observed}) - 2.5 \log C.
\end{eqnarray}
This model is not as formidable as it looks.  $\left(\Delta X, \Delta
Y\right) = (0,0)$ is equivalent to $(X,Y) = (425, 425)$ --- a point near
the center of the unvignetted area on each of the WFPC2 CCDs --- while
$\Delta X = \pm 1$, $\Delta Y = \pm 1$ refer to points near the edges of
the photometrically reliable field.  The peculiar combination of natural
and base-10 logarithms and exponents used here is not entirely whimsical;
for me, at least, it facilitates performing order-of-magnitude estimates
in the head.  For instance, the constant $c_7e^{c_1}$ approximately
represents the fraction of charge lost (expressed in percent) from a star
image containing 10,000 electrons near the effective center of an image
that has a sky surface brightness of 10$\,$e$^-$/pixel.  The constants $\pm
e^{c_1}$ and $\pm e^{c_4}$ represent the degree to which such a star would
lose more or less charge if measured near the edge of the field rather
than near the center as a result of, respectively, the $Y$ and $X$ ramps.
For star images containing either 100,000 or 1,000 electrons, these
corrections would differ by $e^{\pm c_3}$ and $e^{\pm c_6}$ percent,
respectively.  Likewise, for sky values of 100 or $\ltsim 1\,$e$^-$/pixel
(``$\ltsim$'' because of the softening), the corrections for a
10,000-electron star image would differ by $e^{\pm c_2}$ and $e^{\pm c_5}$
percent.  Encoding the corrections as proportional to $\exp\left[\log(\hbox{\it
CTS})\right]$ and $\exp\left[\log( \hbox{\it BKG})\right]$ ensures that
they will go asymptotically --- although not necessarily quickly --- to
zero as $\hbox{\it CTS} \rightarrow \infty$ or $\hbox{\it BKG} \rightarrow
\infty$, as long as $c_2$, $c_3$, $c_5$, and $c_6$ are all less than or
equal to zero.  At the other extreme, the softening of the sky brightness
ensures that the correction approaches a constant value for any given star
in the limit as the background flux becomes negligible.  No corresponding
numerical protection in the limit of negligibly small stellar flux is
required, since the apparent magnitude of a statistically insignificant
star is usually of little astronomical interest.

A copy of the computer program CCDSTD (Stetson 1993) was modified to yield
numerical values and uncertainties for the seven model constants
$c_1,\ldots,c_7$, as well as the zero points of the transformation from
instrumental to standard magnitudes.  As before, the 16 separate zero
points $A_V$ and $A_I$ were computed for the four chips and the two gain
settings (eight values for $A_V$ and eight for $A_I$), based on the cold
(camera temperature = --88\deg$\,$C) data only.  The same gain~=~7$\,$e$^-$/DN
zero points were assumed to apply for the \ocen\ data and both short and
long exposures of \ngc{2419}.  At the same time, the cold data for all four
chips, both filters, both gain settings, and both clusters were used to
determine a single, unique set of values $c_1,\ldots,c_7$.  The numerical
results of this exercise were the zero points listed in Table~3a
\margin{Table~3} and
\newpage
\begin{eqnarray}
c_1 & = & +0.629 \pm 0.021;\nonumber \\
c_2 & = & -0.558 \pm 0.020;\nonumber \\
c_3 & = & -0.593 \pm 0.014;\nonumber \\
c_4 & = & +0.704 \pm 0.027;\nonumber \\
c_5 & = & +0.027 \pm 0.032;\nonumber \\
c_6 & = & -0.359 \pm 0.041;\nonumber \\
c_7 & = & +1.191 \pm 0.018.\nonumber 
\end{eqnarray}
All of these terms $c_i$ appear to be highly significant, with the sole
exception of the sky dependence of the $X$-coordinate ramp ($c_5$): for a
ten-fold change in the sky brightness (\eg, from 10$\,$e$^-$/pixel to
either 1 or 100$\,$e$^-$/pixel) the required correction changes by only
0.03$\pm$0.03\% at the right and left edges of the field, which abundantly
justifies the neglect of this term by WH97.  However, a dependence of the
$Y$ ramp on stellar flux ($c_3$), which WH97 also omitted, appears to be
strongly supported by these data.

According to this model, a star image consisting of 10,000$\,$e$^-$ near
the center of an image with a sky brightness of 10$\,$e$^-$/pixel loses
approximately $c_7e^{c_1} = 1.191e^{+0.629} = 2.2$\% of its electrons; at
$Y=50$ and $Y=800$ it loses $0.191e^{+0.629} = 0.4$\% and
$2.191^{-0.378} = 4.1$\% yielding, in this case, a ramp of about 3.8\%
per 750~pixels or 4.0\% per 800 pixels.  In the hypothetical row $Y=0$ the
same star would lose $\left[\left({-425\over375}\right)
+1.191\right]e^{+0.629} = 0.1$\%, so a position-independent contribution to
the total correction, while apparently statistically significant, is
clearly not very important.  Decreasing the sky brightness by a factor of
ten would increase the $Y$ ramp by a factor $e^{+0.558} \approx 1.75$, to
about 7.0\% per 800 pixels.  For a star 2.5 magnitudes brighter, with
100,000$\,$e$^-$ in its image, the $Y$ ramp would be smaller by a factor
$e^{-0.593} \approx 0.55$, yielding 2.2\% per 800 pixels for
10$\,$e$^-$/pixel in the sky, or 3.9\% per 800 pixels for a sky brightness
of 1$\,$e$^-$/pixel. The numerical correlations between the fitting
residuals and position and magnitude are: 
\begin{eqnarray}
{\partial\delta\over\partial X}\cdot(400\hbox{\ px}) & = & -0.0018\pm 0.0006\,\hbox{mag} \nonumber \\
{\partial\delta\over\partial Y}\cdot(400\hbox{\ px}) & = & -0.0014\pm 0.0006\,\hbox{mag} \nonumber \\
{\partial\delta\over\partial m}\cdot(2.5\hbox{\ mag}) & = & -0.0066\pm 0.0004\,\hbox{mag} \nonumber \\
{\partial^2\delta\over\partial X\partial m}\cdot(400\hbox{\ px})\cdot(2.5\hbox{\ mag}) & = & -0.0007\pm 0.0008\,\hbox{mag} \nonumber \\
{\partial^2\delta\over\partial Y\partial m}\cdot(400\hbox{\ px})\cdot(2.5\hbox{\ mag}) & = & -0.0035\pm 0.0008\,\hbox{mag} \nonumber 
\end{eqnarray}
These are not identically equal to zero because, (a) these linear
correlations do not have the same functional forms as the terms defined in
the least-squares photometric solution, and (b) in the two separate robust
least-squares fits --- one for the model parameters plus independent zero
points for 16 distinct data sets, and one for the correlations between
residuals and positions and magnitudes for all 16 data sets considered
together --- the weighting schemes cannot be perfectly reproduced: the
degrees of freedom are different.  In particular, in fitting the correction
model no allowance was made for a separate trend of photometric error with
magnitude independent of position on the chip, so if such a trend is
present it cannot be perfectly compensated by the model.  As a result, the
single most significant residual correlation is ${\partial
\delta\over\partial m} = (-2.64\pm 0.17)\times 10^{-3}$.  The sign of this
correlation has changed from the uncorrected solutions, indicating that
trend of increasing fractional charge loss with decreasing stellar flux has
been overcorrected, on a global average over the face of the four chips. 
The corrected magnitude scale is now apparently compressed by --0.013~mag
per 5 magnitudes, as compared to the groundbased results.  I believe that
this level of systematic error is small compared to other unavoidable
biases endemic to photometric measurements of very faint stars (see, \eg,
Stetson \& Harris 1988, \S V(b)), and it is unlikely that a much more
satisfactory calibration can be derived from the present data. Certainly,
in view of the crowding conditions in these fields, I am unable to
guarantee that the ground-based photometry is itself linear to this degree
of accuracy. Additional well-established, deep, groundbased photometric
sequences in other star fields targeted by HST, such as the Palomar-type
globular clusters and fields in the Magellanic Clouds, would be most
welcome additions to the study of this problem.  Still, after application
of this set of corrections to the WFPC2 data, the maximum systematic errors
associated with the purely positional variables $Y$ and $X$ are now well
under 1\%, at least averaged over the present data set.  

The secular trend of the photometric residuals with time that remains
after applying these corrections is 
$${\partial \delta \over \partial T} = +0.0012 \pm
0.0002~\hbox{mag~yr}^{-1}.$$ This is larger than was estimated in \S3.1 ---
presumably because in the uncorrected data the mean trend with time was
partially masked by scatter resulting from the neglect of the charge-loss
effects --- but over the 3.5 years of available WFPC2 data this still
implies a maximum departure from the mean zero points of order 0.002~mag at
the two ends of the time interval covered.

\subsection{Summary of the cold-camera data}

There are several additional sanity checks which can be derived from these
data.  For instance, if the corrections truly are valid, then the
photometric zero point for a given chip/filter combination should be
independent of exposure time.  We can perform this test with the short and
long exposures of \ngc{2419}; I omit the \ocen\ exposures from
consideration here because this eliminates confusion produced by any
possible differences between the ground-based photometric systems for the
two clusters.  The four chips and two filters provide us with eight
independent measures of the (short -- long) zero-point difference (data
listed in Tables 1b, 2b, 3b).  Taking the unweighted mean of the eight
(short -- long) zero-point differences, I find net values of $+0.0243 \pm
0.0087$ (standard error of the mean; 0.0245 standard deviation of one
difference) based on the reductions with no photometric corrections;
$+0.0048 \pm 0.0086$ (s.e.m.; 0.0243 s.d.) with the WH97 corrections; and
$+0.0030 \pm 0.0079$ (s.e.m.; 0.0224 s.d.) with the new corrections
derived in \S2.3.  In other words, the WH97 corrections largely eliminate
the perceived difference in zero points between short and long exposures,
but the chip-to-chip and filter-to-filter scatter in the
differences is not much reduced by the WH97 corrections.  The newly
derived corrections represent a slight improvement on those of WH97,
in that they reduce the residual (short~--~long) zero-point
difference a bit further, and also slightly reduce the chip-to-chip and
filter-to-filter scatter.  Repeating the experiment with weighted means
(after all, the PC1 zero points are based on roughly one-fourth as many
stars as those for the wide-field chips) we obtain a mean (short -- long)
difference of $+0.0244 \pm 0.0122$ for the uncorrected data; $+0.0049 \pm
0.0123$ with the WH97 corrections; and $+0.0033 \pm 0.0110$ with the new
corrections.

The gain ratio between Electronics Bay 3 and Electronics Bay 4
may differ from one CCD to the next, but we may expect that the
gain will not depend upon the filter with which a given exposure
was made.  If this is the case, then the difference between
the gain = 14 zero point and the gain = 7 zero point should be
the same in $V$ and $I$.  For instance, with no photometric
corrections, for chip 1 the $V$ zero point from the gain = 14
observations of \ocen\ is 1.7673 (Table 1a), and the gain = 7
zero point is 0.9974 (Table 1b; again, this test considers
the gain = 7 observations for \ocen\ alone to eliminate
the effects of any residual differences between the \ocen\ and
\ngc{2419} ground-based photometry), for a gain ratio of 
0.7699~mag.  The $I$-band gain ratio for the same chip is
$2.6667 - 1.9405 = 0.7262$~mag, and the difference between
the $V$ and $I$ gain ratios is 0.0437~mag, rather than the
expected zero.  Each separate calibration produces four values of this
gain-ratio difference, from the four chips.  The unweighted
root-mean-square value of these four differences is:
0.0245~mag for the uncorrected reductions, 0.0258~mag with
the WH97 corrections, and 0.0285~mag with the new corrections.

For each filter, the zero points of the gain = 14 observations should be
the same in the four chips, because the flat-field observations were made
with the gain = 14 electronics and the flat fields were normalized over the
four chips taken together so that surface brightnesses measured in the
different chips would be restored to a common system.  The standard
deviations of the individual chips about the unweighted mean zero points
for the two filters (eight observations, six degrees of freedom) are 
0.0114~mag for the uncorrected reductions, 0.0098~mag with
the WH97 corrections, and 0.0142~mag with the new corrections.

The tests described in this section are quite indirect, and rely on some
untested assumptions for their validity: that gain ratio is independent of
wavelength, for instance.  Therefore these results can be no more than
suggestive.  There is a hint that the WH97 corrections do better than the
new ones when applied to the \ocen\ data alone, as in the gain-ratio test
and the test assuming the equality of the gain = 14 zero points.  The new
corrections seem to do better when the \ngc{2419} observations are added
to the sample, as in the short--long test here, and in the global fitting
residuals discussed in \S\S3.2 and .3.  If the self-consistency tests of
this section have any meaning at all, it is that the errors in the present
data set are not totally random, but rather contain unmodeled systematic
components as well.  For instance, the photometric zero points may vary
erratically with time, or with the history of previous exposures, or with
some other variable not included in the phenomenological model.  If this is
true, then the standard errors quoted at various places in \S3 and in
Tables~1--3 --- which were estimated simply from the root-mean-square
scatter divided by the square root of the number of fitting residuals ---
are merely attractive fictions; the true errors for any given selected
subset of the data may be considerably larger.  It appears that the
photometric zero point for any given chip/filter combination in any given
data set is really externally reliable only to $\gtsim 0.01$~mag,
regardless of the standard errors listed in the tables.  For instance, the
dispersion of zero-point {\it differences\/} in the (short$\,$--$\,$long) test
just performed was $\sim 0.024$~mag, so the typical uncertainty of the
zero-points themselves would by smaller by $\sqrt{2}$, or
$\sim0.017$~mag.  Similarly, the gain-ratio test involved differences of
differences of zero points, with a scatter $\sim 0.026$, implying zero
points individually reliable to $\sim 0.013$~mag, while the third test,
involving zero points directly, implied a typical external uncertainty in
the range 0.010 -- 0.014~mag.

Whitmore (1998) has conducted an analysis of the time variability of the
charge-transfer inefficiency based on seven images of the \ocen\ standard
field obtained with WFPC2 operating at the colder temperature and spanning
the dates 1994 April to 1997 June.  He concluded that the CTE loss has
increased over that period of time, but only for the faintest stars:
those containing from 20 to 50 DN within a 0\Sec2 aperture in very short
exposures at the 14$\,$e$^-$/DN gain setting appeared to lose $22\pm3$\% of
their electrons when clocked out through 800 rows in the detector at the
end of that period, as compared to $3\pm3$\% at the beginning.  A flux of
280 to 700$\,$e$^-$ within 0\Sec2 corresponds, on my instrumental magnitude
scale, to $m \approx 18.5$ at gain~=~7$\,$e$^-$/ADU and 
$m \approx 19.25$ at gain~=~14.  Accordingly, I have selected out the
photometric residuals, {\it after\/} the application of the new corrections
derived in \S3.3, for stars with predicted instrumental magnitudes
fainter than those limits.  For these stars considered separately, I find
\begin{eqnarray}
{\partial\delta\over\partial X}\cdot(400\hbox{\ px}) & = & -0.0026\pm 0.0007\ \hbox{mag} \nonumber \\
{\partial\delta\over\partial Y}\cdot(400\hbox{\ px}) & = & -0.0028\pm 0.0007\ \hbox{mag} \nonumber \\
{\partial \delta\over \partial T} & = & +0.0025 \pm 0.0002\ \hbox{mag\ yr}^{-1} \nonumber 
\end{eqnarray}
(Systematic variations of the residuals with stellar brightness are not
worth investigating in a sample with this limited range of instrumental
magnitude.) For these faint stars, it seems that the new corrections
overcompensate a bit for the mean rate of charge loss with $X$ and $Y$
position on the chip --- the net trends are slightly negative, indicating
that the corrected magnitudes are a little too bright at positive values
of $\Delta X$ and $\Delta Y$ --- but the maximum
error is $\ll 1$\% and the r.m.s.\ error averaged over the face of the
chip is $\sim 0.0016\,$mag in both $X$ and $Y$.  The residuals for these
very faint star images do show a systematic positive trend with time, as
found by Whitmore:  the mean residual increases at a rate of
0.0025$\,$mag~yr$^{-1}$, for a total net difference of 0.009$\,$mag over the
3.5~years spanned by the current data set for a faint star located near
$(X,Y) = (425,425)$ or, equivalently, for the average of many stars
scattered uniformly over the detector field. Both the overcorrection of the
$X$ and $Y$ ramps and the apparent secular increase of the charge loss with
time are much smaller than the random Poisson noise for any star image
containing 280--700$\,$e$^-$.  Indeed, the random observational errors for
star images this faint are almost always larger than 0.05$\,$mag when all
error contributors are considered.  That is why these stars are essentially
absent from Fig.~1 where, for reasons of clarity, only the more precise
measurements have been plotted.  The unmodeled systematic trends of these
faint-star residuals are also much smaller than the systematic statistical
effects discussed, \eg, by Stetson \& Harris (1988).  Still, while these
remaining systematic effects are unimportant on a star-by-star basis, the
reader should be aware that ensemble average magnitudes for large numbers
of very faint stars observed in short exposures may be subject to these
additional (relatively minor) sources of systematic error.

Whitmore further states that for star images containing at least
2800$\,$e$^-$ within 0\Sec2 (radius) the rate of increase of charge loss
with time is negligible, even in exposures with very low background
levels.  It is probably safe to assume that the secular evolution of the
charge losses will be doubly negligible in images that contain significant
background levels, and that one generic set of charge-loss corrections
will be reasonably valid over the 3--3.5 years spanned by existing
studies.  The results presented here certainly seem to bear that
conclusion out.

\section{Warm Data}

WFPC2 was operated at a camera temperature of --76\deg~C from installation
in 1993 December until 1994 April 23. Between 1994 January 11 and April 21
the \ocen tauri standard field was observed a number of times in F555W
employing both electronics bays, but observations in F814W are only
available for the gain = 14 electronics.  Determining the mean photometric
zero points in the usual way, under the assumption of no charge loss, we
arrive at the fitting residuals shown in Figure~2.\margin{Fig.~2}  There
are comparatively few observations from the short period of time when the
camera was operated at the higher temperature (14,521 residuals are shown
in Fig.~2), and the data do not span the chips nearly as well as in the
more extensive cold data.  It is clear that there is a considerable
charge-loss effect in the $Y$ coordinate, but any systematic trends with
$X$ and magnitude are too small to be seen at this scale.

I have attempted to determine charge-loss coefficients for the warm
data as I did for the cold.  I quickly found that the solution
for the fitting parameter $c_4$ rapidly diverged to $-\infty$,
indicating that any charge loss in the serial register is too small
to measure even in a statistical sense.  In order to get a self-consistent
solution, therefore, it was necessary to simplify the problem.
Accordingly, I effectively turned off the ramp correction in $X$
by assigning $c_4$ a value of --20, and I also eliminated the
possibility of a position-independent charge loss by setting
$c_7$ identically equal to ${425\over375} = 1.1333$.  The standard
robust solution for the remaining three sigificant terms yielded
\begin{eqnarray}
c_1 & = & +1.271 \pm 0.035, \nonumber \\
c_2 & = & -0.449 \pm 0.040, \nonumber \\
c_3 & = & -0.254 \pm 0.011, \nonumber 
\end{eqnarray}
plus the zero points listed in Table~4.\margin{Table~4}
In a frame with a mean background of 10$\,$e$^-$/pixel, the total
amount of charge lost from a 10,000 electron star at $Y=800$
($\Delta Y = 1$) is $2.1333\,e^{\hbox{\small 1.271}} = 7.6$\%, as compared
to $2.191\,e^{\hbox{\small 0.629}} = 4.1$\% for the cold data.  For the
warm data, the variation of the charge loss with background level
appears marginally less than for the cold data, but the dependence on
the brightness of the star is about twice as strong.

\section{Conclusion}

For \ocen, a total of 152 WFPC2 exposures in the F555W filter and 159
exposures in F814W spread over 3.5 years of observations have now been
calibrated to a common photometric system.  In contrast, Walker's (1994)
groundbased study of \ocen\ was based on only 12 $V$ and 10 $I$ frames
obtained over the course of six nights in 1985, using an early RCA CCD.
By combining the HST data with Walker's it should be possible to get an
improved standard sequence in \ocen, ``improved'' at least in the sense of
being internally more precise even if not externally more accurate.
Figure 3(a) \margin{Fig.~3} is a (\vmi, $V$) color-magnitude diagram for
Walker's stars in \ocen, while Fig.~3(b) shows the same stars, but with
the present WFPC2 photometry --- corrected for charge-transfer
inefficiency and transformed to the standard $V,I$ system using the
precepts of \S\S2.3 and 3 and the zero points of Tables 3 and 4 ---
averaged together with Walker's.  Given the large body of HST data, it is
also possible to define new standard-sequence stars in \ocen.
Accordingly, a list of minimally crowded detections in the \ocen\ standard
field was derived from the output of the ALLFRAME reductions.  From among
these relatively isolated stars, a total of 1,318 were selected which had
been measured at least 10 times in each filter, had standard errors of the
mean no larger than 0.03$\,$mag in each of $V$ and $I$, and gave no
evidence of instrinsic variability in excess of 0.03$\,$mag,
root-mean-square, averaged over the two filters.  These are shown in
Fig.~3(c).  (The figure 1,318 includes stars listed by Walker (1994);
1,116 of them are new ones without Walker, Harris, or Woolley
identification numbers as tabulated by Walker.)

Figures~4(a) \margin{Fig.~4} and (b) show, respectively, the ground-based
photometry for \ngc{2419}, and the \hst-based photometry averaged together
with the ground-based data for the same stars.  Since in \ngc{2419} I had
already selected out a large sample of minimimally crowded stars to define
the ground-based standard sequence in the first place, there is no need to
pick out additional local standards based on the HST data alone.

There is an advantage in re-deriving final sets of zero points based upon
these improved, homogenized standard sequences in the two clusters.  With
most of the stars having been measured on more than one of the four chips
during the course of the observations, as well as having been measured in
both long and short exposures (in the case of \ngc{2419}), with the
gain~=~7 and gain~=~14$\,$e$^-$/DN electronics (in the case of \ocen),
and with the camera operating at both the warmer and colder temperatures
(\ocen), any residual systematic errors remaining in the individual
data subsets will be diluted when all are averaged together.  Add to these
considerations the fact that the total number of standards has now been
increased from 534 to 1,714 while the random errors have been
reduced from a median value of 0.017$\,$mag per star to a median of
0.007$\,$mag per star, and it follows that the derivation of new zero
points based on the improved standard sequence will yield photometry
that is more homogeneous across the various data sets, even though the
reanalysis cannot lead to photometry that is more absolutely correct on
the average over all data sets.  The final zero points that result from
calibration of the CTE-corrected data with the augmented list of standards
are given in Table~5.\margin{Table~5}  Again, the reader should note
that the formal standard errors listed in the table are outrageously
optimistic; true uncertainties of order 1\% or slightly more are far
more realistic.  Only additional observations will improve this situation.

Application of the present CTE corrections to other WFPC2 data is
straightforward.  Provided that the reader has stellar positions referred
to the natural coordinate system of each chip ($1 \leq X,Y \leq 800$) and
is careful to use the known gain ratio of the observation to convert the
instrumental magnitude and some estimate of the diffuse sky brightness in
the image to units of electrons, then it is merely necessary to add nine
lines of code, corresponding to Eqs.~(1)--(9), to whatever software the
reader is now using.  If, in addition, the reader happens to be following
the Key Project precepts on the standardization of the raw images (\eg,
the images have been multiplied by 4.0 so that they can be stored as short
integers without significant loss of precision; see Stetson \etal\ 1998),
then the zero points of Table~5 can be used to convert the corrected
half-arcsecond aperture magnitudes to the standard system.  If the reader
does not use the Key Project precepts, Stetson \etal\ (1998) list the
steps necessary to relate these zero points to those of Holtzman
\etal\ (1995b).  The external reliability of photometry on this system is
still probably of order 0.01$\,$mag, since the present zero points
represent a compromise among gain 7 and gain 14, warm- and cold-camera,
and short- and long-exposure data sets, each of which has zero points
individually uncertain at the 0.01--0.02$\,$mag level.  Internally,
however, the various chips should yield photometry consistent at a level
somewhat better than this.

\acknowledgments

I am extremely grateful to Dr.~Jeremy~Mould and the entire staff of
Mount Stromlo and Siding Spring Observatories for the financial
support and hospitality during the time when much of this work was
being done.  Additional financial support from the Jet Propulsion
Laboratory of the California Institute of Technology is also
very much appreciated.

\newpage

\centerline{\bf FIGURE CAPTIONS}

\figcaption[fig1.eps]{Fitting residuals of individual stellar measurements
from photometric solutions where no corrections for charge losses have
been applied.  For clarity, only those measurements with individual
uncertainties smaller than 0.05$\,$mag have been plotted, although poorer
measurements were included in the numerical analysis.  Residuals are
plotted against (top)~$X$-coordinate position in the natural reference
frames of the individual chips, (middle)~$Y$-coordinate position, and
(bottom)~the instrumental magnitude of the star image, which is related to
the number of photoelectrons it contains.  Only observations obtained at
the colder camera temperature (--88\deg$\,$C) are included.}

\figcaption[fig2.eps]{As in Fig.~1, except that only observations obtained
at the warmer camera temperature (--76\deg$\,$C) are included.}

\figcaption[fig3.eps]{($V$,\vmi) color-magnitude diagrams for the
WFPC2 standard field in \ocen tauri.  (a) Walker's (1994) ground-based
measurements.  (b)~Calibrated WFPC2 photometry averaged with Walker's
for the same stars as in (a).  (c)~Calibrated WFPC2 photometry for
an enlarged sample of stars, as described in the text.}

\figcaption[Fig4.eps]{($V$,\vmi) color-magnitude diagrams for the
WFPC2 field in \ngc{2419}.  (a)~Ground-based photometry obtained
by the author from archival data.  (b)~Calibrated WFPC2 photometry
averaged together with the ground-based data for the same stars as
in (a).}

\newpage

\small
\baselineskip 5 pt
\def\P{\phantom{$\,\pm\,$}}
\begin{deluxetable}{rcccc}
\tablecaption{Chip Zero Points, Assuming No Corrections
\label{tbl:zpt1}}
\tablewidth{0pt}
\tablehead{
\colhead{Data set} & \colhead{PC1} & \colhead{WF2} & \colhead{WF3} &
\colhead{WF4} }
\startdata
&\multicolumn{4}{c}{a.\ Combined solutions} \nl
\tablevspace{8 pt}
cold, gain = 14, $V$ & 1.7673 $\pm$ 0.0029 & 1.7515 $\pm$ 0.0010 & 1.7824 $\pm$ 0.0008 & 1.7593 $\pm$ 0.0012 \nl
cold, gain = 14, $I$ & 2.6667 \P 0.0021 & 2.6534 \P 0.0007 & 2.6695 \P 0.0010 & 2.6513 \P 0.0010 \nl
\tablevspace{8 pt}
cold, gain = 7, $V$ & 0.9957 \P 0.0023 & 0.9860 \P 0.0016 & 0.9922 \P 0.0015 & 0.9897 \P 0.0016 \nl
cold, gain = 7, $I$ & 1.9728 \P 0.0021 & 1.8963 \P 0.0012 & 1.9240 \P 0.0012 & 1.9473 \P 0.0011 \nl
\tablevspace{8 pt}
&\multicolumn{4}{c}{b.\ Separate solutions, gain = 7}\nl
\tablevspace{8 pt}
$\omega$~Cen, $V$ & 0.9974 \P 0.0030 & 1.0041 \P 0.0024 & 1.0056 \P 0.0029 & 0.9965 \P 0.0024 \nl 
$\omega$~Cen, $I$ & 1.9405 \P 0.0031 & 1.9029 \P 0.0017 & 1.9035 \P 0.0025 & 1.9134 \P 0.0016 \nl
\tablevspace{8 pt}
\ngc{2419}, short, $V$ & 0.9979 \P 0.0044 & 0.9709 \P 0.0026 & 1.0050 \P 0.0020 & 0.9899 \P 0.0023 \nl
\ngc{2419}, short, $I$ & 1.9981 \P 0.0026 & 1.8929 \P 0.0015 & 1.9367 \P 0.0014 & 1.9594 \P 0.0013 \nl
\tablevspace{8 pt}
\ngc{2419}, long, $V$ & 0.9938 \P 0.0061 & 0.9874 \P 0.0032 & 0.9591 \P 0.0031 & 0.9743 \P 0.0037 \nl
\ngc{2419}, long, $I$ & 1.9468 \P 0.0039 & 1.8826 \P 0.0038 & 1.8900 \P 0.0023 & 1.9224 \P 0.0023 \nl
\enddata
\end{deluxetable}

\newpage

\begin{deluxetable}{rcccc}
\tablecaption{Chip Zero Points, Assuming WH97 Corrections
\label{tbl:zpt1}}
\tablewidth{0pt}
\tablehead{
\colhead{Data set} & \colhead{PC1} & \colhead{WF2} & \colhead{WF3} &
\colhead{WF4} }
\startdata
&\multicolumn{4}{c}{a.\ Combined solutions} \nl
\tablevspace{8 pt}
cold, gain = 14, $V$ & 1.7264 $\pm$ 0.0026 & 1.7254 $\pm$ 0.0009 & 1.7517 $\pm$ 0.0008 & 1.7326 $\pm$ 0.0014 \nl
cold, gain = 14, $I$ & 2.6217 \P 0.0021 & 2.6263 \P 0.0007 & 2.6355 \P 0.0009 & 2.6219 \P 0.0011 \nl
\tablevspace{8 pt}
cold, gain = 7, $V$ & 0.9610 \P 0.0023 & 0.9594 \P 0.0016 & 0.9666 \P 0.0014 & 0.9649 \P 0.0016 \nl
cold, gain = 7, $I$ & 1.9345 \P 0.0018 & 1.8640 \P 0.0010 & 1.8942 \P 0.0010 & 1.9195 \P 0.0010 \nl
\tablevspace{8 pt}
&\multicolumn{4}{c}{b.\ Separate solutions, gain = 7}\nl
\tablevspace{8 pt}
$\omega$~Cen, $V$ & 0.9571 \P 0.0032 & 0.9765 \P 0.0023 & 0.9777 \P 0.0028 & 0.9680 \P 0.0026 \nl 
$\omega$~Cen, $I$ & 1.8967 \P 0.0033 & 1.8715 \P 0.0014 & 1.8744 \P 0.0026 & 1.8795 \P 0.0019 \nl
\tablevspace{8 pt}
\ngc{2419}, short, $V$ & 0.9584 \P 0.0038 & 0.9379 \P 0.0023 & 0.9744 \P 0.0018 & 0.9617 \P 0.0023 \nl
\ngc{2419}, short, $I$ & 1.9539 \P 0.0021 & 1.8554 \P 0.0012 & 1.9024 \P 0.0012 & 1.9285 \P 0.0011 \nl
\tablevspace{8 pt}
\ngc{2419}, long, $V$ & 0.9734 \P 0.0058 & 0.9712 \P 0.0032 & 0.9433 \P 0.0031 & 0.9672 \P 0.0037 \nl
\ngc{2419}, long, $I$ & 1.9267 \P 0.0036 & 1.8668 \P 0.0038 & 1.8729 \P 0.0024 & 1.9128 \P 0.0023 \nl
\enddata
\end{deluxetable}

\newpage

\begin{deluxetable}{rcccc}
\tablecaption{Chip Zero Points, Assuming New Corrections
\label{tbl:zpt1}}
\tablewidth{0pt}
\tablehead{
\colhead{Data set} & \colhead{PC1} & \colhead{WF2} & \colhead{WF3} &
\colhead{WF4} }
\startdata
&\multicolumn{4}{c}{a.\ Combined solutions} \nl
\tablevspace{8 pt}
cold, gain = 14, $V$ & 1.7168 $\pm$ 0.0025 & 1.7300 $\pm$ 0.0009 & 1.7556 $\pm$ 0.0008 & 1.7365 $\pm$ 0.0015 \nl
cold, gain = 14, $I$ & 2.6096 \P 0.0022 & 2.6315 \P 0.0007 & 2.6368 \P 0.0008 & 2.6275 \P 0.0012 \nl
\tablevspace{8 pt}
cold, gain = 7, $V$ & 0.9601 \P 0.0026 & 0.9678 \P 0.0016 & 0.9750 \P 0.0014 & 0.9732 \P 0.0016 \nl
cold, gain = 7, $I$ & 1.9343 \P 0.0019 & 1.8720 \P 0.0010 & 1.8991 \P 0.0009 & 1.9252 \P 0.0009 \nl
\tablevspace{8 pt}
&\multicolumn{4}{c}{b.\ Separate solutions, gain = 7}\nl
\tablevspace{8 pt}
$\omega$~Cen, $V$ & 0.9478 \P 0.0039 & 0.9850 \P 0.0024 & 0.9861 \P 0.0028 & 0.9760 \P 0.0029 \nl 
$\omega$~Cen, $I$ & 1.8891 \P 0.0036 & 1.8776 \P 0.0013 & 1.8844 \P 0.0024 & 1.8897 \P 0.0022 \nl
\tablevspace{8 pt}
\ngc{2419}, short, $V$ & 0.9621 \P 0.0039 & 0.9466 \P 0.0023 & 0.9824 \P 0.0017 & 0.9709 \P 0.0022 \nl
\ngc{2419}, short, $I$ & 1.9542 \P 0.0019 & 1.8646 \P 0.0012 & 1.9059 \P 0.0011 & 1.9337 \P 0.0010 \nl
\tablevspace{8 pt}
\ngc{2419}, long, $V$ & 0.9818 \P 0.0062 & 0.9791 \P 0.0032 & 0.9525 \P 0.0030 & 0.9720 \P 0.0036 \nl
\ngc{2419}, long, $I$ & 1.9359 \P 0.0041 & 1.8743 \P 0.0038 & 1.8815 \P 0.0023 & 1.9187 \P 0.0023 \nl
\enddata
\end{deluxetable}

\newpage

\begin{deluxetable}{rcccc}
\tablecaption{Chip Zero Points, Assuming New Corrections
\label{tbl:zpt1}}
\tablewidth{0pt}
\tablehead{
\colhead{Data set} & \colhead{PC1} & \colhead{WF2} & \colhead{WF3} &
\colhead{WF4} }
\startdata
warm, gain = 14, $V$ & 1.7805 $\pm$ 0.0019 & 1.7381 $\pm$ 0.0011 & 1.7682 $\pm$ 0.0008 & 1.7612 $\pm$ 0.0009 \nl
warm, gain = 14, $I$ & 2.5957 \P 0.0037 & 2.5852 \P 0.0010 & 2.6099 \P 0.0008 & 2.6244 \P 0.0011 \nl
\tablevspace{8 pt}
warm, gain = 7, $V$ & 1.0656 \P 0.0047 & 0.9833 \P 0.0043 & 1.0298 \P 0.0025 & 1.0311 \P 0.0037 \nl
\enddata
\end{deluxetable}

\newpage

\begin{deluxetable}{rcccc}
\tablecaption{Improved Chip Zero Points, Assuming New Corrections
\label{tbl:zpt1}}
\tablewidth{0pt}
\tablehead{
\colhead{Data set} & \colhead{PC1} & \colhead{WF2} & \colhead{WF3} &
\colhead{WF4} }
\startdata
cold, gain = 14, $V$ & 1.7222 $\pm$ 0.0010 & 1.7282 $\pm$ 0.0005 & 1.7475 $\pm$ 0.0004 & 1.7285 $\pm$ 0.0005 \nl
cold, gain = 14, $I$ & 2.6123 \P 0.0008 & 2.6313 \P 0.0003 & 2.6355 \P 0.0004 & 2.6353 \P 0.0003 \nl
\tablevspace{8 pt}
cold, gain = 7, $V$ & 0.9743 \P 0.0012 & 0.9767 \P 0.0005 & 0.9820 \P 0.0004 & 0.9830 \P 0.0005 \nl
cold, gain = 7, $I$ & 1.9088 \P 0.0012 & 1.8792 \P 0.0004 & 1.8957 \P 0.0004 & 1.9140 \P 0.0004 \nl
\tablevspace{8 pt}
warm, gain = 14, $V$ & 1.7836 \P 0.0007 & 1.7441 \P 0.0004 & 1.7683 \P 0.0004 & 1.7602 \P 0.0005 \nl
warm, gain = 14, $I$ & 2.5923 \P 0.0013 & 2.5941 \P 0.0004 & 2.6122 \P 0.0004 & 2.6159 \P 0.0005 \nl
\tablevspace{8 pt}
warm, gain = 7, $V$ & 1.0643 \P 0.0034 & 0.9997 \P 0.0016 & 1.0356 \P 0.0017 & 1.0264 \P 0.0024 \nl
\enddata
\end{deluxetable}
\end{document}